\documentclass[]{pasj00}
\draft 
\begin{document}
\SetRunningHead{Author(s) in page-head}{Running Head}
\Received{2002/02/14}
\Accepted{2002/08/07}

\title{A new empirical method of estimation of the far-infrared flux of galaxies}
\author{Hirohisa \textsc{Nagata}, Hiroshi \textsc{Shibai}}
\affil{Graduate School of Science, Nagoya University, Furo-cho, 
Chikusa-ku, Nagoya, 464--8602}
\email{hnagata@u.phys.nagoya-u.ac.jp}
\author{Tsutomu T. \textsc{Takeuchi}\thanks{A Research Fellow of the 
Japan Society of the Promotion of Science (JSPS).}}
\affil{National Astronomical Observatory, Mitaka, Tokyo 181--8588}
\and
\author{Takashi \textsc{Onaka}}
\affil{Department of Astronomy, Graduate School of Science, 
The University of Tokyo, 7--3--1 Hongo,\\ Bunkyo-ku, Tokyo 113--0033}
\KeyWords{ISM: dust, extinction---galaxies: ISM---galaxies: spiral}
\maketitle

\begin{abstract}
We propose a new empirical method to estimate the total far-infrared
flux of galaxies from the spectral energy 
distribution (SED) at wavelengths $\lambda \leq 100\; \micron$. 
It is difficult to derive the total far-infrared luminosity from 
only the IRAS data, though it is one of the most important properties 
of galaxies.
Observations by Infrared Telescope in Space (IRTS) indicate that
 the SED of the diffuse emission from the Galactic 
plane in this wavelength region can be derived from the $60\; \micron$
 to $100\; \micron$ color. 
This empirical SED relation was improved in order to obtain a 
better fit to the Galactic plane data for 
$I_\nu(60\;\micron)/I_\nu(100\;\micron) > 0.6$,
 and applied to 96 IRAS galaxies for which ISOPHOT and KAO data are 
available at $\lambda > 100\; \micron$. As a result, the empirical 
relation describes well the far-infrared (FIR) SED for a majority of 
the galaxies. Additionally, the total FIR flux for $\lambda \geq 40\; \micron$ 
is derived from the flux densities at 60 and $100\; \micron$ by using 
this model. 
For the 96 IRAS galaxies, the uncertainty in the total far-infrared 
flux of the present method is 26~\%. The present method is more accurate 
than the previous one widely used to derive the total infrared flux from 
the IRAS 60 and $100\; \micron$ data.
\end{abstract}

\section{Introduction}\label{sec:intro}

In the interstellar space of galaxies, there exist solid particles of 
sub-micron sizes, so called ``interstellar dust grains.''
Earlier work (e.g., \cite{beichman}; \cite{soifer1}) has revealed 
that the far-infrared emission from galaxies is often dominated by thermal 
emission from such dust grains.
Therefore, the total luminosity in the far-infrared from the interstellar dust grains is the most important indicator of star-formation 
activity or activity of hidden AGNs. 
Moreover, the far-infrared color is the most direct indicator of the strength of interstellar radiation field 
(ISRF). 

Over the past two decades, many studies have been made to understand 
the properties of infrared emission from interstellar dust grains. 
In order to explain the observed spectral energy distribution (SED) 
of the Galactic plane, 
\citet{desert} proposed a three-component model that consists of 
large grains (LGs), 
very small grains (VSGs), and polycyclic aromatic hydrocarbons (PAHs). 
The large grain is the classical interstellar dust grain of sub-micron size, 
which is thought to be in equilibrium with the interstellar radiation field. 
Bulk of the emission from the large grains is radiated in the far-infrared 
and sub-millimeter wavelength regions. The very small grain is
the dust grain heated transiently up to $\mbox{a few}$ hundred~K by 
the incidence of a single UV photon. A number of dust models 
(e.g., \cite{draine2}) have shown that both stochastically heated 
VSGs with radii of less than 10~nm and very large molecules with 
up to a few hundred atoms could be the source of the excess emission shorter than 100 $\micron$. 
\citet{dwek} showed that the mean infrared SED of the Galactic plane was
 fitted with a composite model of silicate and graphite grains. 
If we take the silicate dust grain to correspond to the large 
grain, and the graphite to the very small grain, these two models have 
essentially the same approach to reproduce the observed SED of the Galactic 
plane. 

By using the Infrared Telescope 
in Space (IRTS) and the Infrared Astronomical Satellite (IRAS) data,
\citet{okumura2} and \citet{shibai} proposed another approach. 
They assumed that the SED between 100 and 200 $\micron$ is dominated by a 
single-component dust with an emissivity index of 2 at a nearly constant 
temperature. 
They showed that the observed SED can successfully be fitted with the 
single component model at a constant temperature, which means that 
the strength of the interstellar radiation field (ISRF) is nearly a constant,
except for the high-temperature regions around distinct H\emissiontype{II} 
regions. However, the diffuse infrared emission at wavelengths $\lambda$ 
shorter than $100\; \micron$ cannot be explained by the thermal 
emission from the large grains, and shows the intensity is 
significantly larger. \citet{okumura2} found that the infrared 
SED varies with the strength 
of interstellar radiation field, and the $12\mbox{--}60\; \micron$ SED 
is strongly correlated with that at $\lambda \geq 100\; \micron$. 
\citet{shibai} obtained the same result from the Galactic plane data 
($|b| < 5$ degree) of COBE/DIRBE. 
These results suggest that the temperature of the large grain is a 
key parameter to characterize the Galactic diffuse infrared emission.

Extending the Galactic plane result to external galaxies, 
one would like to use the largest available database i.e., the 
IRAS survey (e.g., \cite{soifer1}). 
But because of the lack of photometric band at $\lambda > 100\; \micron$, 
the IRAS observations alone are not adequate to characterize the emission from 
dust cooler than 30~K. 
The ISO and the Submillimeter Common-User Bolometer Array (SCUBA) data 
have shown that the far-infrared emission from late-type galaxies has 
a substantial contribution from large grains whose temperature is 
about 15--25 K (\cite{alton}; \cite{haas}; \cite{dunne}). 
If sub-millimeter fluxes are included, the total far-infrared 
emission increases by a factor of 2--3 with respect to previous 
estimates based on IRAS data alone (e.g., \cite{helou2}). 
The derived large grain temperatures of galaxies are significantly lower 
than the previous results estimated from the IRAS $60\; \micron$ and $100\;
\micron$ flux densities (e.g., \cite{devereux}). Additionally, 
\citet{devereux} and \citet{alton} also suggested that the dust masses
 of spiral galaxies derived from sub-millimeter/ISO fluxes are ten times larger than those derived only from the IRAS data. Therefore, the SED at $\lambda > 100\;\micron$ is crucial to estimate the thermal emission from large grains 
correctly, and to understand the physical environments in galaxies. 

To derive the total infrared luminosities of IRAS galaxies, \citet{xu2} 
have proposed an empirical relation between the ratio of the integrated
far-infrared flux (40 -- 120 $\micron$) to the total luminosities 
(8--1000~$\micron$) and the 60-to-100 $\micron$ flux densities ratio 
from 13 late type galaxies. Here, we chose another
approach based on the dust emission properties to 
derive the total infrared luminosities of IRAS galaxies. We derived the large grain temperatures and the total radiation energies from external galaxies in the far-infrared region by using the relationship between the COBE/DIRBE 60 $\micron$, 100 $\micron$ and 140 $\micron$ intensities in the Galactic plane derived by \citet{shibai}.
We also adopted a new empirical model. The present method improves not
only the previous evaluation of the far-infrared total luminosities of 
IRAS galaxies, but also provides the key parameters to characterize 
galaxies, e.g., the temperature and the dust mass of galaxies.

In this paper, we describe our empirical method to derive the total
infrared luminosities of galaxies. The dataset and analysis are explained 
in Section 2. The results and discussion are presented in Section 3 and Section 4, respectively. We summarize the results and present conclusions in Section 5. 
\section{Sample and data analysis}\label{sec:data}

\subsection{COBE/DIRBE data}
We used the COBE/DIRBE ZSMA (Zodi-Subtracted Mission Average) Maps at 60 $\micron$, 100 $\micron$, and $140\; \micron$ for the diffuse Galactic 
emission. These data are available at the Astrophysics Data Facility in
the NASA Goddard Space Flight Center (http://space.gsfc.nasa.gov/astro/cobe).
The reason why we used the three intensities is that \citet{shibai} and \citet{okumura2} have found a correlation between the intensities at 60~$\micron$ and that at $\lambda
\geq$100~$\micron$ (100, 140 and 155~$\micron$ bands).
We selected the area of the Galactic plane at $|b| < 5^\circ $, but
the regions where the 60 $\micron$ intensities, $I_{\nu}(60\; \micron) < 3\; 
\rm MJy\, sr^{-1}$ were omitted in this study to avoid 
the residual contamination after the subtraction of the Zodiacal light.
A total of the omitted areas is 8.5~\% of all the selected area. 
The analysis method used here is the same as that of \citet{shibai}. 
We assumed that the SED at $\lambda \geq 100\; \micron$ is fitted by a 
single temperature Planck function with an emissivity index of
2 and applied color-corrections to the intensities at 
$100\; \micron$ and $140\; \micron$. These correction factors are listed in COBE/DIRBE Explanatory supplement version 2.3 (e.g., Appendix B). 
Then, the corrected intensity $I_\nu ^{c}(\lambda)$ is described simply as
\begin{equation}
 I_{\nu}^{c}(\lambda)= I_{\nu}(\lambda)/K_{\nu}(T_{\rm LG}) = \tau_{100}\cdot\left(\frac{\nu}{3\rm{\;THz}}\right)^{2}\cdot B_{\nu}(T_{\rm LG}) \,\hspace{5mm} [\rm{W \,m^{-2}\, Hz^{-1}\, sr^{-1}}],
\label{plgp}
\end{equation}
where $K_{\nu}(T_{\rm LG})$, $\tau_{100}$, $T_{\rm LG}$, and 
$B_{\nu}(T_{\rm LG})$ are the color-correction factor for DIRBE dataset, the optical depth at $100\; \micron$, the dust temperature derived from the intensities at $100\; \micron$ and $140\; \micron$ (``LG'' denotes the large grain that is in equilibrium with the ISRF), and the Planck function at temperature $T_{\rm LG}$, respectively. 
The dust temperature and the optical thickness are uniquely determined, 
given the $100\; \micron$ and $140\; \micron$ intensities. 
For the temperature range of interest, the color-corrections were less than 8~\%. 

\subsection{Nearby Galaxies}
We selected 97 nearby galaxies observed by ISOPHOT and 
by the Kuiper Airborne Observatory (KAO) from 
\citet{telesco}, \citet{klaas}, \citet{alton}, \citet{haas},
\citet{siebenmorgen}, \citet{calzetti}, \citet{contursi}, 
 \citet{perez}, and \citet{tuffs}, while the 
IRAS $60\; \micron$ and $100\; \micron$ data were taken from 
IRAS Faint Source Catalog (Version 2), Cataloged Galaxies and 
Quasars Observed in the IRAS Survey (Version 2), \citet{edelson}, 
 \citet{soifer2}, \citet{xu}, \citet{alton} and \citet{contursi}.  
These published galaxy samples by IRAS, ISOPHOT, and KAO are summarized 
in Table~1. The sample galaxies collected by us are most of the 
published data observed at $\lambda > 100~\micron$. 
The present sample contains various types of galaxies, Seyfert
galaxies, starbursts, and normal galaxies, etc.
The morphological types of the sample galaxies 
are as follows: 77 spirals, 
11 irregulars, and 9 galaxies with other/unknown type. 
\citet{alton} suggested the possibility that their original ISOPHOT data 
have a calibration error of +30~\%. 
They gave two reasons for their suggestion: 
 1) The integrated flux at 200 $\micron$ derived from the
observation of \citet{engargiola} is smaller than the ISO flux for
NGC 6946. 2) They compared the observed 200 $\micron$ background with
that extrapolated from 100 $\micron$, using the Galactic spectrum at
high latitude from \citet{reach}, and then obtained the same result as 1).
In this paper, we have applied this correction to their data. 
We used the 150 $\micron$ data for three galaxies (NGC 3516, UGC 12138, Mrk
533) of \citet{perez} and the 170 $\micron$ data for two galaxies (Mrk 57, CGCG 160139) of 
\citet{contursi} because the $200\; \micron$ flux densities are significantly lower than the $90\mbox{--}150\; \micron$ flux densities in the former case 
(See Figure~2 of \cite{perez}) and than the 120 and 150 $\micron$ flux 
densities in the latter case (See Figure~6 of \cite{contursi}). 
The data listed in Table~1 are catalog values and need to be applied a color-correction. The correction factors are given in the ISO handbook Volume V (e.g., Appendix C) (\cite{laureijs})
and the Infrared Astronomical Satellite Explanatory Supplement 
(e.g., Section VI) (\cite{iras}). Since the color-correction factor was not available for the KAO 140 $\micron$ flux density of M~82, it was not applied. For the SED at $\lambda \geq 100\; \micron$, we assumed that the dust emissivity index of each sample galaxy is proportional to ${\lambda^{-2}}$, and fitted the two photometric data beyond $100\; \micron$ with the 
following equation:
\begin{equation}\label{plg}
  F_{\nu}^{c}(\lambda)=F_{\nu}(\lambda)/K_{\nu}(T_{\rm LG})=\Omega\cdot\tau_{100}\cdot\left(\frac{\nu}{3\rm{\;THz}}\right)^{2}\cdot B_{\nu}(T_{\rm LG}) \,\hspace{5mm} [\rm{W \,m^{-2}\, Hz^{-1}}],
\end{equation}
where $F_{\nu}(\lambda)$, $F_{\nu}^{c}(\lambda)$, $\Omega$, and $K_{\nu}(T_{\rm LG})$ are the catalog flux density in Table~1, the color corrected flux density, the solid angle of the object, and the color-correction factor of IRAS or ISOPHOT, respectively. The other parameters are the same as that in Equation (\ref{plgp}). Here we note that $\Omega$ and $\tau_{100}$ cannot be determined independently. For the temperature range of interest, the corrections to the catalog flux densities were found to be less than 14 \%.
\citet{haas} and \citet{contursi} have derived temperatures of 
galaxies from the data at $\lambda \geq 100\; \micron$ under the same assumption. We confirmed that the values for $T_{\rm LG}$ derived by us are consistent with the temperatures derived by them. 

It is noted that we derived the dust temperature, $T_{\rm LG}$, from color corrected intensities/flux densities by using Equation (1) or (2), whereas use catalogued intensities/flux densities without the color-correction for the other parameters given in subsequent sections i.e., the dust temperature, $T_{\rm LG2}$, and the intensities/fluxes ratio of Figure 2 and 3.
\section{Results}

\subsection{The Galactic plane}\label{sec:gp}

Figure~\ref{fig1} shows the relation between the intensity 
$\nu I_{\nu} (60\;\micron)$ and the dust temperature, $T_{\rm LG}$ 
derived from the intensities $I_{\nu}(100\; \micron)$ and $I_{\nu}(140\;  \micron)$. 
Intensity $\nu I_{\nu}(60\; \micron)$ is normalized by the optical depth at 
$100\; \micron$. The intensity $\nu I_{\nu}(60\; \micron)$ shows a clear power-law dependence. This correlation cannot be fitted by a single temperature Planck function with an emissivity index of 2. We call the simple model of $\lambda^{-2}B_{\nu}(T_{\rm LG})$ ``Model~1." 
\citet{okumura2} and \citet{shibai} found the following correlation for the Galactic plane far-infrared emission at 60, 100, and 140 or 155 $\micron$.
\begin{equation}\label{okumura}
\frac{\nu I_{\nu} (60\;\micron)}{I_{\rm FIR}}\propto G, 
\end{equation}
where $G$ is the strength of the ISRF given by Equation (\ref{habing}), and $I_{\rm FIR}$ is the total far-infrared intensity given by Equation (\ref{okumura2}). They adopted Draine and Lee model (\cite{draine1}) which assumes 
the ratio of average dust absorption efficiency in the ultraviolet 
to the infrared is 700 in thermal equilibrium. Then, the 
ISRF strength, $G$ can be written as follows: 
\begin{eqnarray}\label{habing}
G &\,=\,& 1.84\times10^{-7}\cdot G_{0}\cdot T_{\rm LG}^{6} \,=\, 2.99\times 10^{-13}\cdot T_{\rm LG}^{6}     \hspace{20mm} {\rm [W \, m^{-2}]},  \\
{\rm and} \nonumber \\ 
G_{0}&=&1.6\times 10^{-6}\, \,\hspace{10mm} [\rm{W \,m^{-2}}], \nonumber 
\end{eqnarray}
where $G_{0}$ is the strength of ISRF in the solar neighborhood (\cite{habing}). 
By Using Equation (\ref{habing}), $I_{\rm FIR}$ is proportional to $G$ as shown by the following relation,
\begin{equation}\label{okumura2}
 I_{\rm FIR}=\int_{0}^{\infty}\tau_{100}\cdot\left(\frac{\nu}{3\rm{\;THz}}\right)^{2}\cdot B_{\nu}(T_{\rm LG})\, d\nu \;\propto\; \tau_{100}\cdot G \,\hspace{10mm} [\rm{W \,m^{-2}\,sr^{-1}}]. 
\end{equation}
From Equations (\ref{okumura}), (\ref{habing}) and (\ref{okumura2}), the following relation
is derived: 
\begin{equation}
\frac{\nu I_{\nu} (60\;\micron)}{\tau_{100}}=1.8\times 10^{-19}\times(T_{\rm LG})^{12} \,\hspace{5mm} [\rm{W \,m^{-2}\,sr^{-1}}].
\label{shibai0}
\end{equation}
The coefficient of Equation (\ref{shibai0}) is slightly modified from
the original relation derived by \citet{shibai}, so as to get a 
better fit to the data of Figure~1. The relation given
 by Equation (\ref{shibai0}) is called ``Model~2."

Figure~2 shows the relation between the dust temperature, $T_{\rm LG}$, and the ratio, $I_{\nu}(60\;\micron)$/$I_{\nu}(100\; \micron)$. The data points near the brighter end ($I_{\nu}(60\;\micron)$/$I_{\nu}(100\; \micron)>$0.6) show the intensities lower than those described by Equation (\ref{shibai0}) as can be also seen in Figure~1. Here, we present a better fit to the correlation between the ratio, $I_{\nu}(60\;\micron)$/$I_{\nu}(100\; \micron)$, and the dust temperature, $T_{\rm LG}$, plotted in Figure~2 by employing a simple empirical relation as follows:
\begin{equation}\label{est0}
\frac{I_{\nu}(60\; \micron)}{I_{\nu}(100\; \micron)}=\frac{(T_{\rm LG}-13.8)}{11.8}.
\end{equation}
The empirical relation of Equation (\ref{est0}) is called ``Model~3." It is clearly seen
that the new relation can fit the data in Figure \ref{fig2} better than
the previous models at $T_{\rm LG}>17~\rm{K}$. Although this empirical 
fit by Model~3 for $T_{\rm LG}<17~\rm{K}$ is worse than that for Model~2, 
it is a minor flaw for the main 
conclusion because almost all $T_{\rm LG}$s of external galaxies in our
sample are higher than $17~\rm{K}$ (See the next section). 

Equation (\ref{est0}) suggests that the dust temperature can also be 
derived from $I_{\nu}(60\; \micron)$ and $I_{\nu}(100\; \micron)$ in the
Galactic plane. The temperature derived from the 60 and $100\; \micron$
bands is called ``$T_{\rm LG2}$" to distinguish it from $T_{\rm LG}$. The
derived temperature, $T_{\rm LG2}$, is given by the following relation:
\begin{equation}\label{est}
T_{\rm LG2} = 11.8\times\frac{I_{\nu}(60\; \micron)}{I_{\nu}(100\; \micron)}+13.8 \,\hspace{5mm} [\rm{K}].
\end{equation}
The uncertainty in the estimation of $T_{\rm LG2}$ by Equation~(\ref{est}) is approximately 10 $\%$ for $T_{\rm LG}>\; 16$ K.

\subsection{Nearby Galaxies}
The dust temperatures derived for the sample galaxies are 
listed in Table~2. 
The temperatures of the sample galaxies range from 16~K to 41~K with an 
average of 24~K. 
For comparison of $T_{\rm LG}$ and the $60\;\micron$ to $100\;\micron$
color of nearby galaxies with those of the Galaxy, the mean Galactic plane
intensities at 60, 100 and $140\; \micron$ were used to derive the color and
the temperature, $T_{\rm LG}$, of the Galactic plane. The derived temperature of the Galaxy is 
$T_{\rm LG}\sim 17~\rm{K}$, and the intensity ratio is $I_{\nu}(60\;\micron)/I_{\nu}(100\; \micron)\sim 0.3$. 
A majority of the sample galaxies are warmer than the Galactic plane.
Figure~\ref{fig3} shows the relation between the ratio,
 $F_{\nu}(60\; \micron)$/$F_{\nu}(100\; \micron)$, and the dust 
temperature, $T_{\rm LG}$. The curve of Model~1 in Figure~3 is changed from that in Figure~2 by the difference in the color-correction factors between IRAS and DIRBE. 
The difference in the curve is less than 14~\% for $T_{\rm LG} \,>$ 30~K, but from 14 to 40~\% for 20~K $<\,T_{\rm LG} \,<$ 30~K. The reason why the difference is relatively large for $T_{\rm LG} \,<$ 30~K is that the difference in the color-correction factors at 60~$\micron$ becomes sensitive to that in the detailed spectral shapes between IRAS and DIRBE when the thermal emission in 60~$\micron$ band is in the Wien region. For Model~2 and Model~3, the differences in the two curves are small for 16~K $<\,T_{\rm LG} \,<$ 50~K i.e., less than 8.5~\%. Therefore, the same curves as in Figure~2 are given in Figure~3.  
There is a trend that $F_{\nu}(60\; \micron)$/$F_{\nu}(100\; \micron)$ 
increases with the dust temperature.
Most of the data points of galaxies are located between the curves of 
Model~1 and
Model~3 in Figure~\ref{fig3}. This trend of the sample galaxies is
similar to that of the Galactic plane given by Equation (\ref{est0}). 
When $T_{\rm LG} >$ 25~K, the data points approach the
curve of Model~1. The
data point of the spiral galaxy, Mrk 993, is significantly apart from 
the curve given
by Equation (\ref{est0}). The flux of Mrk 993 may have a large
calibration error since the flux density of IRAS 100 $\micron$ of Mrk
993 is more than twice the ISOPHOT 90--200 $\micron$ flux
densities. Mrk 993 is excluded in the following analysis. This
exclusion does not at all affect the main conclusion of this paper.

Next, we derived the dust temperature, $T_{\rm LG2}$, from $F_{\nu}(60\;
\micron)$/$F_{\nu}(100\; \micron)$ by using Equation (\ref{est}). 
The derived dust temperature, $T_{\rm LG2}$, is listed in Table~2. 
As seen in Figure~3, for a majority of galaxies, the derived temperatures by Equation (\ref{plg}) are larger
than $T_{\rm LG2}$ and less than the temperature derived by
Model~1. The difference between $T_{\rm LG}$ and $T_{\rm LG2}$ is
1--43 \%. The root mean square of the difference ($T_{\rm LG2}$ - $T_{\rm LG}$) is 19 \%. As the temperatures become higher, the differences
become larger. For $T_{\rm LG} <$ 25~K, the root mean square of the
difference is 14 \%, while for $T_{\rm LG} >$ 25~K, it is 26~\%. 

\section{Discussion}

\subsection{Relation between the ratio $F_{\nu}(60\; \micron)/F_{\nu}
(100\; \micron)$ and $T_{\rm LG}$}
As shown in Figure~\ref{fig3}, the relation between the ratio 
$F_\nu(60\; \micron)/F_\nu(100\; \micron)$ and $T_{\rm LG}$ determined from 
the spectrum at $\lambda \geq 100\; \micron$ for the present nearby galaxy 
sample is similar to that of the Galactic plane.
At $\lambda \geq 100\; \micron$, the diffuse infrared emission from the 
Galactic plane is considered to be dominated by thermal emission from 
large grains, which are in equilibrium with the ISRF (e.g., \cite{mathis3}). 
Therefore, the temperature of the large grains depends on the strength of the 
ambient radiation field. 
In other words, the far-infrared color at longer wavelengths
is the best indicator of the strength of the ISRF in the observed line
of sight.

\begin{figure}
\hspace{2.2cm}
 \FigureFile(100mm,120mm){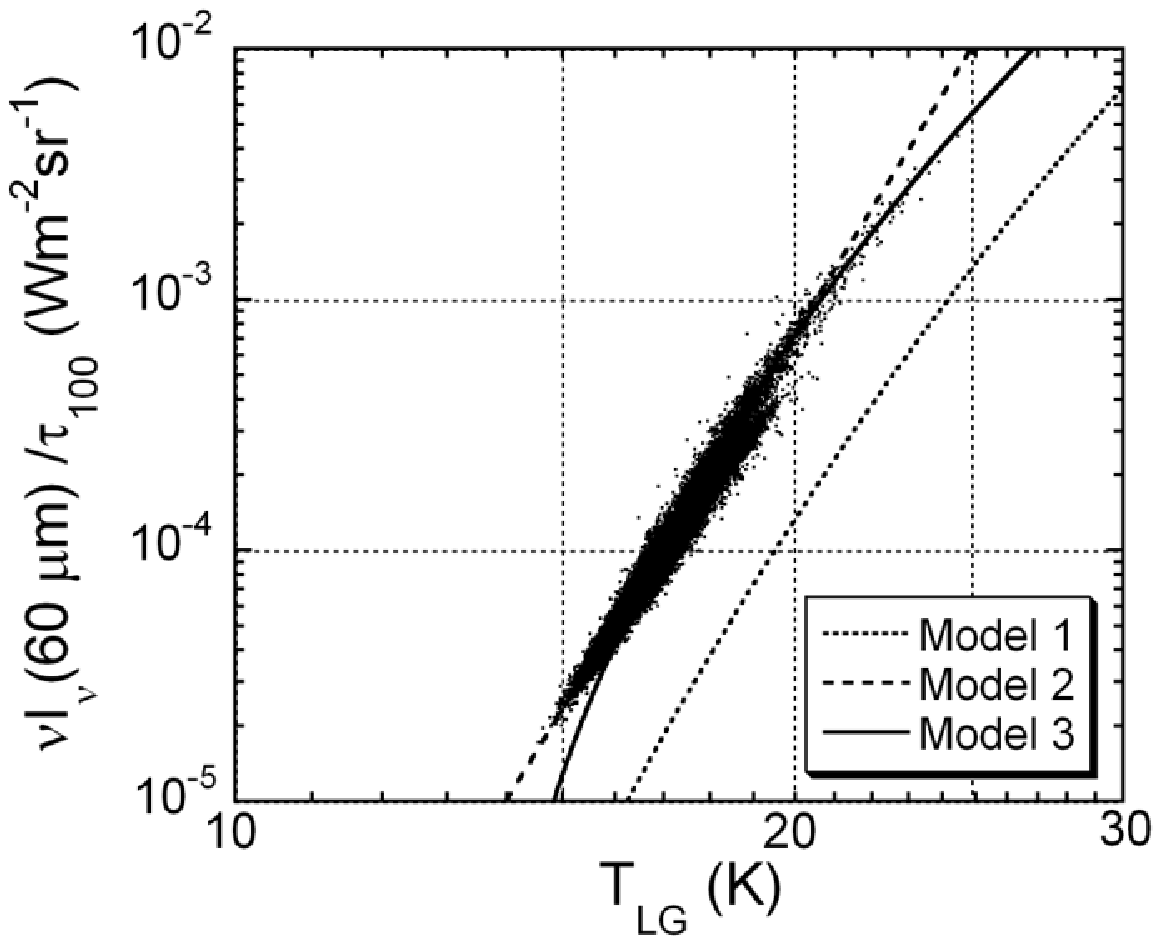}
\caption{Plot of the DIRBE 60~$\micron$ intensity normalized by the 100~$\micron$ optical depth, $\nu I_{\nu}(60\; \micron)/\tau_{100}$, vs. the dust temperature, $T_{\rm LG}$, derived from the DIRBE 100~$\micron$ and 140~$\micron$ intensities for the Galactic plane in the region within 5 degree from the Galactic equator. 
The three curves in this figure indicate the relations between $\nu I_{\nu}(60\; \micron)/\tau_{100}$ and $T_{\rm LG}$ derived from the three SED models, respectively. The dotted curve (Model~1) indicates that derived from a single temperature Planck function with an emissivity index of 2. The dashed curve (Model~2) indicates that derived from the relation described as Equation~(\ref{okumura}). The solid curve (Model~3) indicates that given by the present empirical relation described as Equation (\ref{est0}). 
} 
\label{fig1}
\end{figure}

\begin{figure}
 \begin{center}
\FigureFile(100mm,120mm){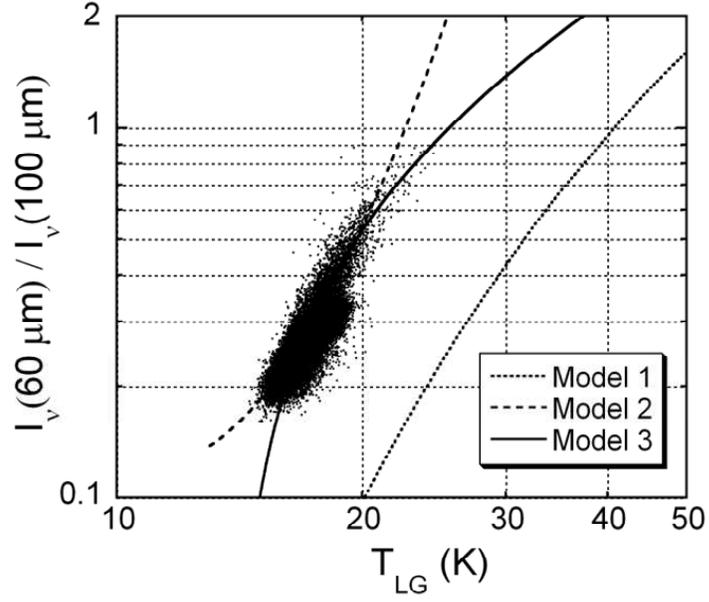}
 \end{center}
\caption{Plot of the ratio of DIRBE 60~$\micron$ to 100~$\micron$ intensities, 
$I_{\nu}(60\; \micron)/I_{\nu}(100\; \micron)$, vs. the dust temperature,
 $T_{\rm LG}$, derived from the DIRBE 100~$\micron$ and 140~$\micron$ intensities for the Galactic plane in the region within 5 degree from the Galactic equator. The three curves in this figure indicate the relations between $I_{\nu}(60\; \micron)/I_{\nu}(100\; \micron)$ and $T_{\rm LG}$ derived from the same three SED models as in Figure~\ref{fig1}.}\label{fig2}
\end{figure}

\begin{figure}
 \begin{center}
\FigureFile(100mm,120mm){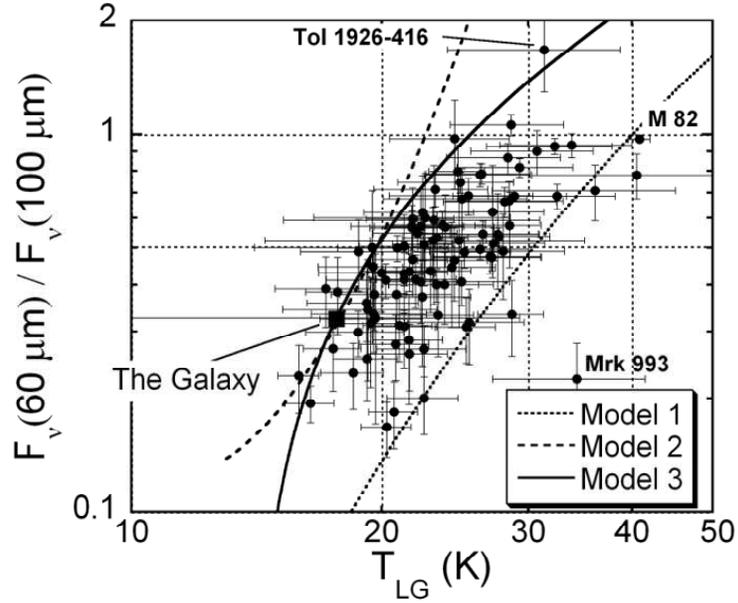}
 \end{center}
\caption{Plot of the ratio of IRAS 60~$\micron$ to 100~$\micron$ flux densities, $F_{\nu}(60\; \micron)/F_{\nu}(100\; \micron)$, vs. the dust temperature, $T_{\rm LG}$, derived from the IRAS flux density at 100~$\micron$ and the ISOPHOT (KAO) flux density at $\lambda \,>\,100~\micron$ for the galaxy sample listed in Table~1. The data points of solid circle are galaxies of Table~1, and the error bars of the sample galaxies are the 1$\sigma$ uncertainty. The data point of solid square 
symbol is derived from the three bands ($60\; \micron$, $100\; \micron$, 
$140\; \micron$) dataset which are the mean intensities of the Galactic plane in the region within 5 degrees from the Galactic equator. The three curves in this figure indicate the relations between $F_{\nu}(60\; \micron)/F_{\nu}(100\; \micron)$ and $T_{\rm LG}$ derived from the same three SED models as in Figure~\ref{fig1}.}
\label{fig3}
\end{figure}

On the other hand, the diffuse infrared emission at $\lambda < 100\; 
\micron$ cannot be explained by the thermal emission from large grains. 
\citet{desert} showed that the $60\; \micron$ emission comes mainly from 
very small grains (VSGs). 
Thus, the $60\; \micron$ emission and the emission at 
$\lambda \geq 100\; \micron$ come from different dust components. 
Figure~\ref{fig1} indicates that the $60\; \micron$ intensity 
normalized by the optical depth at $100\; \micron$ has 
a tight correlation with 
the dust temperature determined by the spectrum at $\lambda \geq 100\;
\micron$, though it consists of intensities in various
directions of the Galactic plane. This correlation is well fitted by Equation (\ref{est0}).
Since Equation~(\ref{est0}) does not have a clear physical basis, we call 
this a new empirical SED model for the $60\; \micron$ emission. 
This implies that the mixing ratio of VSGs with large grains is nearly 
constant throughout the Galactic plane. 
Otherwise, we would see much larger scatter in Figures~\ref{fig1} 
and~\ref{fig2}. \citet{sodroski} and \citet{shibai} showed that 
the far-infrared emission from the Galactic plane is dominated by 
emission from large 
grains at $T_{\rm LG} \sim 17$~K in equilibrium with the ambient ISRF 
and not by warm dust grains existing in or near discrete H\emissiontype{II} 
regions. 
Therefore, most of the far-infrared radiation of the Galactic plane does 
not come from discrete H\emissiontype{II} regions, but from interstellar 
dust grains diffusely distributed with a uniform temperature heated 
by the general ISRF. 
This interpretation is supported by the result that a large fraction of 
the UV photons from early-type stars in the Galactic plane are not absorbed and form a 
general ISRF as suggested by \citet{mezger}.

We now apply the new empirical SED model to nearby galaxies. 
As seen in Figures~\ref{fig2} and \ref{fig3}, these extragalactic objects 
show the same relation as the Galactic plane. 
This directly indicates that, for a majority of the galaxies sampled here, 
the far-infrared emission comes predominantly from the diffuse component 
heated by the general ISRF like the Galactic plane, 
and that the far-infrared SED of the galaxies have the same properties as 
that of the Galactic plane. Thus, the 
$F_\nu(60\; \micron)/F_\nu(100\;  \micron)$ ratio can well be reproduced 
from the SED at $\lambda \geq  100\; \micron$, and the infrared
 SEDs (longer than $100\; \micron$) can also be reproduced from
 $F_\nu(60\; \micron)/F_\nu(100\;  \micron)$. 

\citet{dale2} suggested that the infrared SED in the 
$6\mbox{--}25\; \micron$ region of normal galaxies can be reconstructed
from the ratio of $F_\nu(60\; \micron)/F_\nu(100\;  \micron)$ by an
analysis of $6\mbox{--}100\; \micron$ SED of 69 normal galaxies. Taking
their result, the whole infrared SED longer than $6~\micron$ can be
described only by the $F_\nu(60\; \micron)/F_\nu(100\;  \micron)$ 
ratio. Recently, Dale and Helou (2002) developed a new SED model of
galaxies in the wavelength range from $3~\micron$ to 20~cm. This model
is based on an empirical characteristic of spectra of galaxies by assuming a
simple relation of the dust emissivity to the strength of the
interstellar radiation field without physical explanation. In contrast,
the present work is unique because it is derived based on observations of 
the infrared SED of the Galactic plane.

The uniformity in the infrared SED of galaxies suggests that the dust 
properties are common in galaxies as well as the effective 
dust-formation processes. Moreover, as described in Section 3, no
systematic difference in the infrared SED is seen in different activity/morphological types 
of galaxies. \citet{perez1} concluded from their ISOPHOT observations of 
CfA Seyfert sample that not only the far-infrared SED of starburst galaxies 
but also the SED of the Seyfert galaxies can be interpreted in terms of the 
thermal emission from star-forming regions. \citet{negishi} concluded by 
spectroscopic observations of nearby galaxies with ISO/LWS that 
there is no difference in far-infrared line ratios between starbursts 
and AGNs. Accordingly, we propose that the far-infrared 
SEDs are commonly described by the present empirical model 
regardless of galaxy type and morphology.

As seen in Figure \ref{fig3}, several galaxies with 
$T_{\rm LG} > 25$ K can be fitted better by a single temperature 
Planck function with the emissivity index of 2 (Model~1) 
than by Model~3 of Equation (\ref{est0}). 
This fact indicates that the $60\; \micron$ emission is close to 
the thermal emission from warm large grains in intense interstellar 
radiation field. It may be the case, if we consider that the contribution 
of discrete H\emissiontype{II} regions whose emission peaks at
wavelength shorter than 100 $\micron$ to the total infrared SED becomes 
larger in galaxies of 
higher $T_{\rm LG}$.
For example, the luminous nearby infrared galaxy, M~82, can be well fitted 
by a single temperature Planck function with the emissivity index of 2 and  
M~82 lies near the curve of Model~1 as shown in Figure~\ref{fig3}.
On the other hand, Tol 1924$-$416 is located close to the present empirical 
SED model (Equation (\ref{est0})), while the dust temperature, $T_{\rm LG}$, is high, 31.4~K. 
This galaxy is a famous blue compact irregular galaxy, and 
the star-forming regions of the galaxy do not seem to contain much dust.
This is inferred from its low Balmer decrement value H${\alpha}$/H${\beta} 
\sim 3.1$ (\cite{storchi-bergmann}). 
The small dust content of this galaxy allows 
most of the UV/optical photons radiated by the stars formed to escape from 
their star-forming clouds easily. These escaped photons contribute to 
the general interstellar radiation field, and are absorbed by grains 
in the diffuse interstellar space. 

We adopted the value of 2 as the dust emissivity index throughout this
paper. This value was derived from the SED of the Galactic Plane
(\cite{okumura2}; \cite{reach}), and is naturally predicted by the
standard dust grain model (\cite{draine1}). Some galaxies reveal smaller
values of this index in the far-infrared region, e.g., M 82
(\cite{colbert}), Arp 220, Arp 244 and NGC 6240 (\cite{klaas}). However,
the value of this index does not affect the content of the present paper, if it is between 1 and 2. 

\subsection{Application to total infrared luminosity}
\label{sec:apTIE}
\citet{helou2} proposed that the total far-infrared flux in the wavelength
region between 40 $\micron$ and 120 $\micron$, FIR, is given by
\begin{equation}\label{fir}
  \mbox{FIR}=1.26\times 10^{-14}\times(2.58\times F_{\nu}(60\; \micron) +F_{\nu}(100 \; \micron)) \,\hspace{5mm} [\rm{W \, m^{-2}}].
\end{equation} 
 However, as they noted, Equation (\ref{fir}) is valid only when 
the far-infrared emission that peaks between 50 and 100 $\micron$. 
Thus, FIR is limited to
an indicator of total far-infrared emission because it neglects the emission
at shorter and longer wavelengths. To obtain the total far-infrared
integrated flux for 
$\lambda \geq40 \micron$, we define $\mbox{FIR1}$, derived from the three 
photometric bands; the $60\; \micron$ band and two photometric bands longer than $100\; \micron$. $\mbox{FIR1}$ is given as follows,  
\begin{eqnarray}\label{lg}
  \mbox{FIR1}& \approx &\int_{0}^{\infty}\Omega\cdot 
  \tau_{100}\cdot\left(\frac{\nu}{3\rm{\;THz}}\right)^{2}\cdot 
  B_{\nu}(T_{\rm LG})\, d\nu\nonumber\\
 &+& \left(F_{\nu}(60\; \micron) -  B_{\nu}(\lambda = 60~\micron, \,\,T_{\rm LG})\cdot\Omega\cdot\tau _{100}\cdot\left(\frac{60}{100}\right)^{-2}\right)\cdot\Delta \nu_{60},
\end{eqnarray}  
where $B_{\nu}(\lambda = 60~\micron,\,\, T_{\rm LG})$ is the Planck function of the temperature, $T_{\rm LG}$, at $60\; \micron$, and $\Delta \nu_{60}$ is
the frequency width of the IRAS $60\; \micron$ band, which
corresponds to $3.75 \times10^{12}$ Hz. 
The first term shows the total far-infrared emission from large grains, and 
the second term shows the contribution of the excess emission in the 
$60\; \micron$ band. Therefore, FIR1 provides a total infrared 
integrated flux for $\lambda \geq 40\; \micron$. Here, if the integration of the first term is carried out and $\Omega\cdot\tau_{100}$ is replaced with $F_{\nu}^{c}(100 \; \micron)$/$B_{\nu}(\lambda = 60~\micron,\,\, T_{\rm LG})$, then FIR1 can be rewritten by the following relation of $F_{\nu}(60\; \micron)$, $F_{\nu}^{c}(100 \; \micron)$ and $T_{\rm LG}$.
\begin{eqnarray}\label{fir1-2}
  \mbox{FIR1} \approx C_{1}\cdot F_{\nu}(60\; \micron)&+&F_{\nu}^{c}(100 \; \micron)\times({\rm exp}({\textstyle{\frac{143.88}{T_{\rm LG}}}})-1) \nonumber\\
& \times &\left(C_{2}\cdot T_{\rm LG}^{6}-\frac{C_{3}}{{\rm exp}({\textstyle{\frac{239.79}{T_{\rm LG}}}})-1}\right) \,\hspace{13mm} [\rm{10^{-26}\,W \,m^{-2}}],
\end{eqnarray} 
where $C_{1}$, $C_{2}$, and $C_{3}$ are $3.75 \times10^{12}$~Hz, $41.2~\rm{Hz\cdot K^{-6}}$, and $4.82 \times10^{13}$~Hz respectively. Both the flux densities of $F_{\nu}(60\; \micron)$ and $F^{c}_{\nu}(100 \; \micron)$ are in units of Jansky. 
Figure 4 shows the comparison of $\mbox{FIR1}$ with the previous
estimate, FIR. The ratio of FIR to FIR1 sharply decreases with $T_{\rm
LG}$ at $T_{\rm LG} <$ 20~K. As suggested by \citet{helou2}, FIR
underestimates the total far-infrared flux of galaxy at low temperatures
(especially, for $T_{\rm LG} <$ 20~K), although FIR is a good estimator 
at high temperatures (especially, for $T_{\rm LG} >$ 30~K).   

As shown in Section~3, when only the fluxes at 60 and $100\;
\micron$ are available, the dust temperature, $T_{\rm LG2}$, of galaxies can be 
derived from the flux densities at 60 and 100 $\micron$ by using Equation (\ref{est}). Thus, 
the far-infrared integrated flux can be estimated from Equation (\ref{fir1-2}), when we use $T_{\rm LG2}$ and $F_{\nu}(100 \; \micron)$ instead of $T_{\rm LG}$ and $F_{\nu}^{c}(100 \; \micron)$. We denote this estimate as FIR2.
Figure~5 shows the relation between the ratio of FIR2 to FIR1 and
$T_{\rm LG}$. Figure~5 suggests that FIR2 approximates FIR1 well when $T_{\rm LG} <$ 20~K, and that FIR2 estimates FIR1 with the error between about 0~\% and +70~\% when $T_{\rm LG} >$ 20~K. The estimated total fluxes, FIR1 and FIR2, are 
listed in Table~2. For about 70~\% of the sample galaxies, the
differences between FIR1 and FIR2 are less than 25~\%  and less than 30~\% 
for 80~\% of the sample galaxies. The root mean
square of the difference for the entire sample is 26~\%. 
Thus, the total far-infrared integrated flux can be estimated from the 
IRAS Point Source Catalog and Faint Source Catalog by 
FIR2 with an uncertainty of 20--30~\%. We conclude the present method (FIR2) 
is more useful than the previous one (FIR) for 15~K$<T_{\rm LG}<$50~K. 

The total infrared emission is one of the most crucial indicators of 
star-forming activity (\cite{sanders}; \cite{soifer1}). 
Therefore, the present method gives more reliable values for the 
star formation rate of IRAS galaxies that have only the 60 
and $100\; \micron$ flux densities. 
Moreover, other important properties of galaxies, namely, 
dust mass, averaged general interstellar radiation field, and dust 
temperature can be estimated solely from the 60 and $100\; \micron$ 
flux densities by using the present method. 

\section{Conclusion}
We have successfully improved the empirical SED model proposed by 
\citet{okumura2} and \citet{shibai} in order to obtain a better fit to 
the diffuse infrared emission from the Galactic plane. 
This new model is more accurate at higher color temperature.
We applied the model to the infrared SEDs of nearby galaxies 
observed by IRAS and ISO.
\begin{enumerate}
\item    
We found that the majority of the sample galaxies showed a trend similar 
to the Galactic plane, but some galaxies approach a single 
temperature Planck function with the emissivity index of 2 when 
the large grain temperature, $T_{\rm LG}$, is higher than 25 K. 
This fact suggests that for galaxies of $T_{\rm LG}<25$ K
the diffusely distributed dust grains at a single temperature heated by 
the general interstellar radiation field dominate as in the Galactic 
plane, while, for some galaxies of $T_{\rm LG}>25$ K, discrete warm 
components from H\emissiontype{II} regions significantly contribute to 
the averaged SED in the far-infrared region and possibly dominate the
60 $\micron$ and $100\; \micron$ bands over the diffuse component.
\\
\item
The SED at $\lambda >100\; \micron$ can be derived from the 
flux densities at 60 and $100\; \micron$ by the present model.
For 96 IRAS galaxies, the uncertainty in the total far-infrared 
flux for $\lambda \geq 40~\micron$ derived from the present method is 26~\%. 
This method allows us to obtain more accurately the important 
properties of galaxies, such as dust mass, large grain temperature, and 
total infrared flux of IRAS galaxies.
\end{enumerate}

\bigskip
We thank to Drs. K.\ Okumura and T.\ Hirao for their important 
suggestions, and to the IRTS and ASTRO-F members for useful discussion.
We wish to express our gratitude to Dr. T.\ N.\ Rengarajan for his critical
reading of this paper. We thank Dr. Simone Bianchi, the referee, for his careful reading and useful comments that improved this paper very much.
This research has made use of the NASA/IPAC 
Extragalactic Database (NED), which is operated by the Jet Propulsion Laboratory, California Institute of 
Technology, under contract with the National Aeronautics and Space 
Administration.  One of the authors (Tsutomu T. Takeuchi) is financially supported by the JSPS Fellowship.
This research has been supported in part by a Grant-in-Aid for 
the Scientific Research Fund (10147102) of the Ministry of Education, 
Science, Sports and Culture in Japan.

\begin{figure}
 \begin{center}
\FigureFile(100mm,120mm){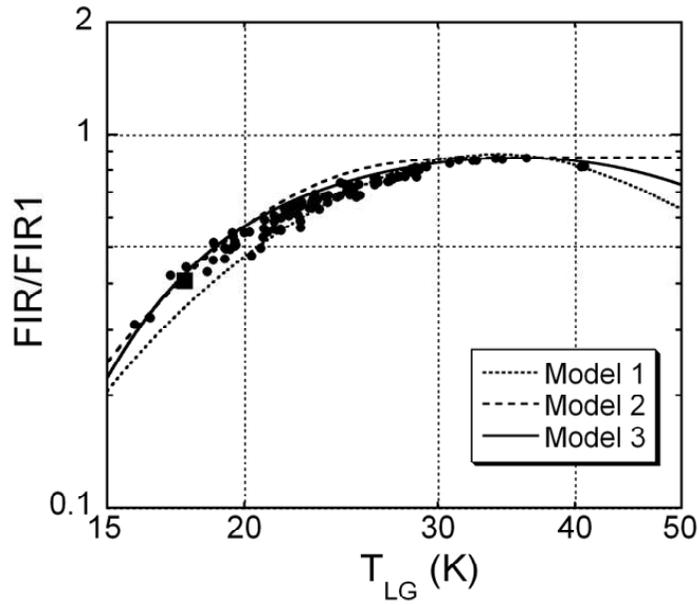}
 \end{center}
\caption{The ratio of FIR/FIR1 vs. $T_{\rm LG}$ for the galaxy sample listed in Table~1. FIR is the total infrared flux in the wavelength region between 40 and 120~$\micron$ given by \citet{helou2}. FIR1 is the total infrared flux for $\lambda\,\geq\,40~\micron$ derived from Equation (\ref{fir1-2}). 
The dust temperature, $T_{\rm LG}$, is the same as in Figure~3. The symbols of data points are the same as in Figure~\ref{fig3}. The three curves in this figure indicate the relations between FIR/FIR1 and $T_{\rm LG}$ derived from the same three SED models as in Figure~\ref{fig1}.
 }\label{fig4} 
\end{figure}

\begin{figure}
 \begin{center}
\FigureFile(100mm,120mm){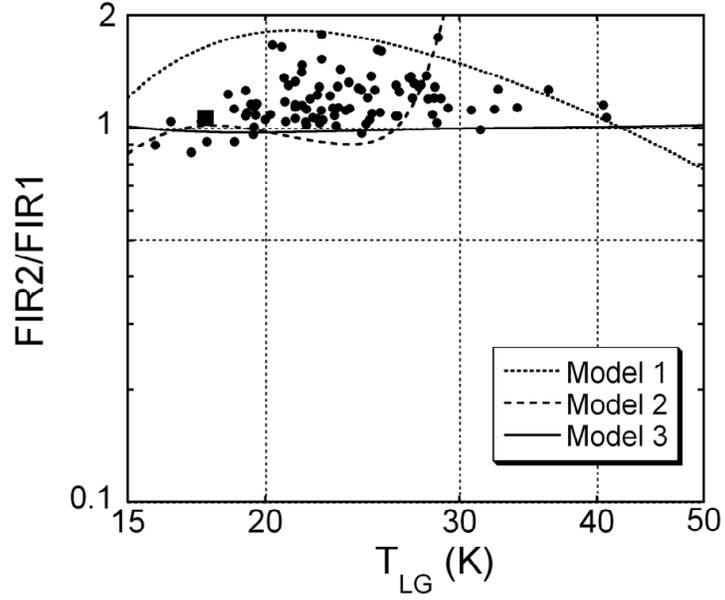}
 \end{center}
\caption{The ratio of FIR2/FIR1 vs. $T_{\rm LG}$ for the galaxy sample listed in Table~1. FIR1 is the total infrared flux for $\lambda\,\geq\,40~\micron$ derived from Equation (\ref{fir1-2}). FIR2 is the estimate of FIR1 derived from the IRAS 60 and 100~$\micron$ flux densities by using Equation~(\ref{est0}). The dust temperature, $T_{\rm LG}$, is the same as in Figure~3. 
 The symbols of data points are the same as in Figure~\ref{fig3}. The three curves in this figure indicate the relations between FIR2/FIR1 and $T_{\rm LG}$ derived from the same three SED models as in Figure~\ref{fig1}.
}\label{fig5}
\end{figure}

\begin{longtable}{lcccccc}
\caption{Galaxy data}\label{table1}
  \hline\hline
Galaxy name &\footnotemark[$*$]$F_{\nu}(60\; \micron)$&\footnotemark[$*$]$F_{\nu}(100\; \micron)$&\footnotemark[$\dagger$]$F_{\nu}(150\; \micron)$&\footnotemark[$\dagger$]$F_{\nu}(170\; \micron)$&\footnotemark[$\dagger$]$F_{\nu}(200\; \micron)$&\footnotemark[$\ddagger$]Refs.\\
 &$(\rm{Jy})$&$(\rm{Jy})$&$(\rm{Jy})$&$(\rm{Jy})$&$(\rm{Jy})$\\
\hline
\endhead
\hline
\endfoot
\hline
 \multicolumn{7}{l}{\hbox to 0pt{\parbox{180mm}{\footnotesize
\bigskip
\footnotemark[$*$]Flux densities at 60 and 100 $\micron$ from IRAS data (which are not color corrected).
The 1 $\sigma$ uncertainties are also reported. 
\\
\footnotemark[$\dagger$]Flux densities at 150, 170 and 200 $\micron$ from ISOPHOT and KAO data (which are also not color corrected). The 1 $\sigma$ uncertainties are also reported. 
\\
\footnotemark[$\ddagger$]References of the sample of
(a)~\citet{telesco}; (b)~\citet{klaas};  (c)~\citet{alton};
(d)~\citet{haas}; (e)~\citet{siebenmorgen}; (f)~\citet{calzetti};  
 (g)~\citet{contursi};  (h)~\citet{perez};  (i)~\citet{tuffs}; 
 (j)~IRAS Faint Source Catalog Version 2; (k)~Cataloged Galaxies and
 Quasars Observed in the IRAS Survey Version 2; (l)~\citet{edelson}; 
(m)~\citet{soifer2}; (n)~\citet{xu}.
\\
\footnotemark[$\S$]The flux density of M 82 in the 150 $\micron$ column of
this table is not that of 150 $\micron$ but of 140 $\micron$.  
}}}
\endlastfoot
NGC 134&$22.4\pm3.4$&$67.0\pm10.1$&-&-&$89\pm27$&(c)\\
M 31&$595\pm28$&$3073\pm316$&-&$7900\pm800$&-&(d), (n)\\
Mrk 555&$4.22\pm0.27$&$8.68\pm0.52$&-&-&$5.3\pm1.6$&(e), (j)\\
IC 1586&$0.96\pm0.08$&$1.69\pm0.19$&$2.11\pm0.52$&-&-&(f), (j)\\
Mrk 993&$0.27\pm0.05$&$1.32\pm0.24$&-&-&$0.39\pm0.12$&(h), (j)\\
NGC 628&$26.3\pm3.9$&$67.5\pm10.1$&-&-&$146\pm44$&(c)\\
NGC 660&$71.4\pm10.7$&$120\pm18$&-&-&$111\pm33$&(c)\\
UGC 1395&$0.47\pm0.05$&$1.24\pm0.17$&-&-&$1.33\pm0.40$&(h), (l)\\
NGC 1068&$176\pm9$&$224\pm9$&-&-&$120.8\pm36.2$&(h), (j)\\
UGC 2936&$4.49\pm0.23$&$10.96\pm0.99$&-&-&$13.0\pm3.9$&(e), (j)\\
UGC 2982&$8.35\pm0.25$&$16.89\pm0.85$&-&-&$9.2\pm2.8$&(e), (j)\\
M 82&$1313\pm1$&$1355\pm1$&\footnotemark[$\S$]$630\pm24$&-&-&(a), (m)\\
Mrk 1243&$0.35\pm0.05$&$0.87\pm0.15$&-&-&$0.62\pm0.17$&(h), (j)\\
NGC 3079&$44.50\pm2.67$&$89.22\pm4.46$&-&-&$95.12\pm28.54$&(h), (j)\\
NGC 3227&$7.83\pm0.39$&$17.59\pm0.88$&-&-&$23.43\pm7.03$&(h), (j)\\
UGC 6100&$0.63\pm0.06$&$1.53\pm0.20$&-&-&$1.38\pm0.41$&(h), (l)\\
NGC 3516&$1.76\pm0.07$&$2.26\pm0.29$&$1.11\pm0.33$&-&-&(h), (j)\\
CGCG 97079&$0.39\pm0.08$&$0.65\pm0.16$&-&-&$0.54\pm0.35$&(g)\\
UGC 6697 (CGCG 97087)&$1.58\pm0.24$&$3.41\pm0.53$&-&-&$2.25\pm0.36$&(g)\\
NGC 3982&$6.57\pm0.53$&$15.23\pm1.22$&-&-&$14.45\pm4.34$&(h), (j)\\
Arp 244&$39.47\pm2.37$&$72.30\pm6.51$&-&-&$64.1\pm19.2$&(b), (j)\\
NGC 4051&$7.13\pm0.93$&$23.92\pm1.20$&-&-&$36.90\pm11.07$&(h), (j)\\
NGC 4151&$6.72\pm0.34$&$8.60\pm0.43$&-&-&$4.69\pm1.41$&(h), (l)\\
NGC 4178 (VCC 66)&$1.54\pm0.25$&$8.39\pm1.01$&-&$11.27\pm0.03$&-&(i), (k)\\
NGC 4192 (VCC 92)&$5.11\pm0.82$&$18.94\pm3.03$&-&$40.29\pm0.50$&-&(i), (k)\\
NGC 4207 (VCC 152)&$3.03\pm0.48$&$7.46\pm1.19$&-&$8.38\pm0.38$&-&(i), (k)\\
NGC 4235&$0.32\pm0.04$&$0.65\pm0.14$&-&-&$0.30\pm0.09$&(h), (j)\\
Mrk 766&$4.03\pm0.28$&$4.66\pm0.28$&-&-&$2.10\pm0.63$&(h), (j)\\
VCC 459&$0.24\pm0.04$&$0.54\pm0.14$&-&$0.50\pm0.03$&-&(i), (j)\\
NGC 4293 (VCC 460)&$4.62\pm0.74$&$10.63\pm1.70$&-&$11.25\pm1.10$&-&(i), (k)\\
NGC 4344 (VCC 655)&$0.43\pm0.07$&$1.88\pm0.23$&-&$5.36\pm0.03$&-&(i), (k)\\
UGC 7470 (VCC 664)&$0.49\pm0.08$&$0.86\pm0.17$&-&$0.97\pm0.03$&-&(i), (k)\\
NGC 4351 (VCC 692)&$0.73\pm0.12$&$1.91\pm0.31$&-&$3.91\pm0.09$&-&(i), (k)\\
NGC 4388&$10.24\pm0.82$&$18.10\pm1.27$&-&-&$16.78\pm5.03$&(h), (j)\\
NGC 4394 (VCC 857)&$1.01\pm0.12$&$4.34\pm0.69$&-&$7.76\pm0.18$&-&(i), (k)\\
NGC 4402 (VCC 873)&$5.80\pm0.93$&$17.47\pm2.80$&-&$17.72\pm0.29$&-&(i), (k)\\
NGC 4413 (VCC 912)&$1.02\pm0.12$&$3.31\pm0.53$&-&$2.82\pm0.10$&-&(i), (k)\\
NGC 4423 (VCC 971)&$0.47\pm0.08$&$1.10\pm0.14$&-&$1.38\pm0.09$&-&(i), (j)\\
NGC 4430 (VCC 1002)&$1.11\pm0.13$&$4.01\pm0.64$&-&$5.30\pm0.11$&-&(i), (k)\\
NGC 4429 (VCC 1003)&$1.54\pm0.25$&$4.62\pm0.74$&-&$3.12\pm0.06$&-&(i), (k)\\
NGC 4438 (VCC 1043)&$4.08\pm0.49$&$10.82\pm1.73$&-&$16.72\pm0.59$&-&(i), (k)\\
NGC 4450 (VCC 1110)&$1.19\pm0.14$&$7.12\pm0.85$&-&$10.01\pm0.17$&-&(i), (k)\\
UGC 7621 (VCC 1189)&$0.23\pm0.06$&$0.73\pm0.14$&-&$1.16\pm0.03$&-&(i), (j)\\
NGC 4477 (VCC 1253)&$0.59\pm0.07$&$1.11\pm0.13$&-&$1.13\pm0.09$&-&(i), (k)\\
NGC 4491 (VCC 1326)&$2.64\pm0.32$&$3.31\pm0.40$&-&$2.98\pm0.31$&-&(i), (k)\\
NGC 4498 (VCC 1379)&$1.20\pm0.14$&$3.86\pm0.62$&-&$4.98\pm0.15$&-&(i), (k)\\
NGC 4502 (VCC 1410)&$0.23\pm0.06$&$0.62\pm0.14$&-&$0.69\pm0.04$&-&(i), (j)\\
UGC 7695 (VCC 1450)&$1.30\pm0.16$&$3.15\pm0.38$&-&$3.96\pm0.13$&-&(i), (k)\\
NGC 4531 (VCC 1552)&$0.35\pm0.05$&$1.75\pm0.25$&-&$1.94\pm0.13$&-&(i), (j)\\
NGC 4532 (VCC 1554)&$8.95\pm1.43$&$15.64\pm2.50$&-&$10.65\pm0.19$&-&(i), (k)\\
UGC 7736 (VCC 1575)&$1.07\pm0.13$&$2.30\pm0.37$&-&$2.74\pm0.14$&-&(i), (k)\\
NGC 4569 (VCC 1690)&$7.21\pm1.15$&$23.36\pm3.74$&-&$29.16\pm0.32$&-&(i), (k)\\
NGC 4579 (VCC 1727)&$4.54\pm0.73$&$17.91\pm2.87$&-&$29.19\pm1.20$&-&(i), (k)\\
NGC 4580 (VCC 1730)&$1.17\pm0.14$&$4.48\pm0.72$&-&$5.46\pm0.55$&-&(i), (k)\\
Mrk 53 (CGCG 160020)&$0.65\pm0.11$&$1.05\pm0.20$&-&-&$0.89\pm0.31$&(g)\\
Mrk 231&$31.99\pm1.60$&$30.29\pm1.21$&-&-&$13.34\pm4.00$&(h), (j)\\
IC 3913 (CGCG 160026)&$0.26\pm0.06$&$0.44\pm0.14$&-&-&$0.34\pm0.27$&(g)\\
NGC 4848 (CGCG 160055)&$1.40\pm0.22$&$2.75\pm0.43$&-&-&$2.31\pm0.35$&(g)\\
Mrk 57 (CGCG 160067)&$0.45\pm0.08$&$0.83\pm0.18$&-&$0.60\pm0.05$&-&(g)\\
CGCG 160086&$0.15\pm0.05$&$0.46\pm0.14$&-&-&$0.60\pm0.60$&(g)\\
UGC 8118 (CGCG 160088)&$0.25\pm0.06$&$0.88\pm0.18$&-&-&$0.84\pm0.15$&(g)\\
IC 4040 (CGCG 160252)&$1.38\pm0.21$&$2.64\pm0.41$&-&-&$2.06\pm0.76$&(g)\\
NGC 4911 (CGCG 160260)&$0.78\pm0.13$&$2.46\pm0.39$&-&-&$1.45\pm0.47$&(g)\\
NGC 4921 (CGCG 160095)&$0.23\pm0.06$&$0.67\pm0.16$&-&-&$0.94\pm0.37$&(g)\\
CGCG 160128&$0.22\pm0.06$&$0.44\pm0.14$&-&-&$0.59\pm0.30$&(g)\\
CGCG 160127&$0.23\pm0.06$&$0.44\pm0.14$&-&-&$0.28\pm0.25$&(g)\\
CGCG 160139&$0.36\pm0.07$&$0.58\pm0.15$&-&$0.43\pm0.05$&-&(g)\\
NGC 5033&$16.45\pm0.1$&$50.81\pm0.1$&-&-&$66.93\pm20.08$&(h), (m)\\
Mrk 66&$0.54\pm0.04$&$0.80\pm0.15$&$0.77\pm0.19$&-&-&(f), (j)\\
NGC 5194&$130\pm19$&$303\pm45$&-&-&$373\pm112$&(c)\\
NGC 5236&$314\pm47$&$624\pm94$&-&-&$622\pm187$&(c)\\
UGC 8621&$1.07\pm0.08$&$2.63\pm0.30$&-&-&$1.64\pm0.49$&(h), (l)\\
NGC 5252&$0.43\pm0.06$&$0.75\pm0.13$&-&-&$0.53\pm0.16$&(h), (l)\\
Mrk 266&$7.34\pm0.44$&$11.07\pm0.78$&-&-&$5.11\pm1.53$&(h), (j)\\
Mrk 461&$0.44\pm0.06$&$0.46\pm0.11$&-&-&$0.30\pm0.09$&(h), (l)\\
Mrk 279&$1.26\pm0.06$&$2.20\pm0.15$&-&-&$1.59\pm0.48$&(h), (j)\\
Mrk 799&$10.41\pm0.42$&$19.47\pm0.80$&-&-&$9.3\pm2.8$&(e), (j)\\
IC 4397&$1.54\pm0.11$&$3.25\pm0.23$&-&-&$1.65\pm0.50$&(h), (j)\\
NGC 5548&$1.07\pm0.09$&$1.61\pm0.16$&-&-&$0.72\pm0.22$&(h), (j)\\
Mrk 817&$2.12\pm0.09$&$2.27\pm0.14$&-&-&$0.69\pm0.21$&(h), (j)\\
NGC 5719&$8.06\pm0.48$&$17.10\pm0.86$&-&-&$8.6\pm2.6$&(e), (j)\\
NGC 5860&$1.64\pm0.08$&$3.02\pm0.18$&$2.63\pm0.65$&-&-&(f), (j)\\
Arp 220&$103.8\pm4.2$&$112.4\pm3.4$&-&-&$37.6\pm11.3$&(b), (j)\\
NGC 6090&$6.66\pm0.26$&$8.94\pm0.45$&$8.67\pm2.15$&-&-&(f), (j)\\
NGC 6104&$0.48\pm0.05$&$1.79\pm0.21$&-&-&$1.50\pm0.45$&(h), (l)\\
NGC 6240&$22.68\pm0.91$&$27.78\pm1.11$&-&-&$11.6\pm3.5$&(b), (j)\\
Tol 1924-416&$1.69\pm0.10$&$1.01\pm0.22$&$0.68\pm0.17$&-&-&(f), (j)\\
NGC 6918&$9.32\pm0.56$&$13.62\pm0.68$&-&-&$4.5\pm1.4$&(e), (j)\\
NGC 6946&$165\pm25$&$338\pm51$&-&-&$520\pm156$&(c)\\
NGC 7331&$42.9\pm6.4$&$120\pm18$&-&-&$170\pm51$&(c)\\
UGC 12138&$0.90\pm0.08$&$1.27\pm0.19$&$0.71\pm0.21$&-&-&(h), (l)\\
Mrk 323&$3.16\pm0.22$&$7.91\pm0.55$&-&-&$6.0\pm1.8$&(e), (j)\\
NGC 7673&$4.91\pm0.34$&$6.89\pm0.48$&$7.60\pm1.90$&-&-&(f), (j)\\
Mrk 533&$5.59\pm0.50$&$8.15\pm0.57$&$7.61\pm2.28$&-&-&(h), (j)\\
Mrk 534&$7.28\pm0.04$&$10.65\pm0.12$&-&-&$4.6\pm1.4$&(e), (m)\\
Mrk 538&$10.36\pm1.24$&$11.51\pm0.69$&-&-&$4.3\pm1.3$&(e), (j)\\
Mrk 332&$4.87\pm0.34$&$9.49\pm0.67$&-&-&$4.7\pm1.4$&(e), (j)\\
\end{longtable}

\begin{longtable}{lccccc}
\caption{Results}\label{table2}
  \hline\hline
   Galaxy name &$T_{\rm LG}$ (K)&$T_{\rm LG2}$ (K)&$\mbox{FIR}$
 (W/m$^{2}$)&$\mbox{FIR1}$ (W/m$^{2}$)&$\mbox{FIR2}$ (W/m$^{2}$)\\ \hline
\endhead
\hline
\endfoot
\hline
 \multicolumn{7}{l}{\hbox to 0pt{\parbox{180mm}{\footnotesize
\bigskip

\footnotemark[$\|$] The results of Mrk 993 are omitted from the analysis in this paper since the flux density at 200~$\micron$ from ISOPHOT 
\\
for the galaxy may have a large calibration error (See section 3-2 in detail). 
}}}
\endlastfoot
NGC 134&$19.6 \pm3.1 $&17.7 &1.57E-12&3.13E-12&3.56E-12\\
M 31&$16.4 \pm1.1 $&16.1 &5.81E-11&1.80E-10&1.87E-10\\
Mrk 555&$25.2 \pm3.4 $&19.5 &2.47E-13&3.49E-13&4.41E-13\\
IC 1586&$21.8 \pm3.4 $&20.5 &5.26E-14&8.34E-14&8.61E-14\\
\footnotemark[$\|$]Mrk 993&$34.3 \pm7.2 $&16.4 &2.63E-14&3.07E-14&7.70E-14\\
NGC 628&$17.1 \pm2.3 $&18.4 &1.71E-12&4.07E-12&3.50E-12\\
NGC 660&$21.8 \pm2.6 $&20.8 &3.83E-12&6.03E-12&6.14E-12\\
UGC 1395&$20.8 \pm3.4 $&18.2 &3.08E-14&5.50E-14&6.45E-14\\
NGC 1068&$26.4 \pm3.8 $&23.1 &8.54E-12&1.12E-11&1.20E-11\\
UGC 2936&$20.2 \pm3.1 $&18.6 &2.84E-13&5.19E-13&5.64E-13\\
UGC 2982&$26.3 \pm3.8 $&19.6 &4.84E-13&6.60E-13&8.58E-13\\
M 82&$40.8 \pm1.2 $&25.2 &5.98E-11&7.31E-11&7.77E-11\\
Mrk 1243&$23.8 \pm3.4 $&18.5 &2.24E-14&3.41E-14&4.51E-14\\
NGC 3079&$20.9 \pm3.2 $&19.7 &2.57E-12&4.36E-12&4.53E-12\\
NGC 3227&$19.5 \pm2.8 $&19.1 &4.76E-13&8.95E-13&8.99E-13\\
UGC 6100&$22.0 \pm2.6 $&18.7 &3.98E-14&6.58E-14&7.87E-14\\
NGC 3516&$40.5 \pm10.2 $&23.0 &8.57E-14&1.05E-13&1.21E-13\\
CGCG 97079&$22.6 \pm6.0 $&20.9 &2.09E-14&3.17E-14&3.33E-14\\
UGC 6697 (CGCG 97087)&$24.5 \pm2.3 $&19.3 &9.43E-14&1.38E-13&1.74E-13\\
NGC 3982&$21.6 \pm2.4 $&18.9 &4.05E-13&6.78E-13&7.78E-13\\
Arp 244&$22.1 \pm2.6 $&20.2 &2.20E-12&3.46E-12&3.68E-12\\
NGC 4051&$18.7 \pm2.5 $&17.3 &5.33E-13&1.16E-12&1.30E-12\\
NGC 4151&$26.3 \pm3.8 $&23.0 &3.27E-13&4.29E-13&4.61E-13\\
NGC 4178 (VCC 66)&$20.7 \pm1.4 $&16.0 &1.56E-13&3.15E-13&5.17E-13\\
NGC 4192 (VCC 92)&$17.5 \pm1.4 $&17.0 &4.05E-13&1.01E-12&1.05E-12\\
NGC 4207 (VCC 152)&$22.3 \pm2.3 $&18.6 &1.93E-13&3.14E-13&3.84E-13\\
NGC 4235&$28.0 \pm5.2 $&19.6 &1.84E-14&2.39E-14&3.28E-14\\
Mrk 766&$28.4 \pm4.4 $&24.0 &1.90E-13&2.36E-13&2.57E-13\\
VCC 459&$24.3 \pm3.1 $&19.0 &1.46E-14&2.16E-14&2.76E-14\\
NGC 4293 (VCC 460)&$22.9 \pm2.7 $&18.9 &2.84E-13&4.46E-13&5.44E-13\\
NGC 4344 (VCC 655)&$15.9 \pm0.9 $&16.5 &3.77E-14&1.22E-13&1.09E-13\\
UGC 7470 (VCC 664)&$22.3 \pm2.7 $&20.5 &2.68E-14&4.15E-14&4.38E-14\\
NGC 4351 (VCC 692)&$17.7 \pm1.5 $&18.3 &4.78E-14&1.08E-13&9.92E-14\\
NGC 4388&$21.8 \pm2.4 $&20.5 &5.60E-13&8.89E-13&9.22E-13\\
NGC 4394 (VCC 857)&$18.5 \pm1.6 $&16.5 &8.75E-14&2.04E-13&2.51E-13\\
NGC 4402 (VCC 873)&$23.4 \pm2.4 $&17.7 &4.09E-13&6.50E-13&9.29E-13\\
NGC 4413 (VCC 912)&$25.3 \pm2.1 $&17.4 &7.49E-14&1.11E-13&1.79E-13\\
NGC 4423 (VCC 971)&$21.3 \pm1.8 $&18.8 &2.91E-14&4.96E-14&5.64E-14\\
NGC 4430 (VCC 1002)&$20.8 \pm2.0 $&17.1 &8.66E-14&1.63E-13&2.22E-13\\
NGC 4429 (VCC 1003)&$28.7 \pm2.6 $&17.7 &1.08E-13&1.41E-13&2.45E-13\\
NGC 4438 (VCC 1043)&$19.6 \pm1.8 $&18.3 &2.69E-13&5.20E-13&5.63E-13\\
NGC 4450 (VCC 1110)&$20.3 \pm1.4 $&15.8 &1.28E-13&2.71E-13&4.50E-13\\
UGC 7621 (VCC 1189)&$19.4 \pm2.1 $&17.5 &1.67E-14&3.39E-14&3.92E-14\\
NGC 4477 (VCC 1253)&$23.3 \pm2.1 $&20.1 &3.32E-14&4.98E-14&5.64E-14\\
NGC 4491 (VCC 1326)&$24.7 \pm1.9 $&23.2 &1.28E-13&1.74E-13&1.78E-13\\
NGC 4498 (VCC 1379)&$21.0 \pm2.0 $&17.5 &8.76E-14&1.60E-13&2.08E-13\\
NGC 4502 (VCC 1410)&$22.4 \pm3.2 $&18.2 &1.53E-14&2.52E-14&3.23E-14\\
UGC 7695 (VCC 1450)&$21.3 \pm1.6 $&18.7 &8.20E-14&1.40E-13&1.62E-13\\
NGC 4531 (VCC 1552)&$22.5 \pm2.2 $&16.2 &3.34E-14&5.94E-14&1.05E-13\\
NGC 4532 (VCC 1554)&$28.5 \pm2.6 $&20.6 &4.88E-13&6.20E-13&7.97E-13\\
UGC 7736 (VCC 1575)&$21.8 \pm2.2 $&19.3 &6.38E-14&1.04E-13&1.17E-13\\
NGC 4569 (VCC 1690)&$21.3 \pm2.0 $&17.4 &5.29E-13&9.50E-13&1.26E-12\\
NGC 4579 (VCC 1727)&$19.2 \pm1.7 $&16.8 &3.73E-13&8.04E-13&1.01E-12\\
NGC 4580 (VCC 1730)&$21.6 \pm2.4 $&16.9 &9.45E-14&1.71E-13&2.51E-13\\
Mrk 53 (CGCG 160020)&$22.4 \pm3.4 $&21.1 &3.44E-14&5.23E-14&5.39E-14\\
Mrk 231&$28.6 \pm4.5 $&26.3 &1.42E-12&1.75E-12&1.80E-12\\
IC 3913 (CGCG 160026)&$23.1 \pm7.9 $&20.8 &1.40E-14&2.09E-14&2.25E-14\\
NGC 4848 (CGCG 160055)&$22.5 \pm1.9 $&19.8 &8.02E-14&1.25E-13&1.40E-13\\
Mrk 57 (CGCG 160067)&$27.6 \pm3.4 $&20.2 &2.51E-14&3.27E-14&4.22E-14\\
CGCG 160086&$19.7 \pm9.6 $&17.6 &1.07E-14&2.12E-14&2.45E-14\\
UGC 8118 (CGCG 160088)&$21.6 \pm2.1 $&17.2 &1.92E-14&3.44E-14&4.84E-14\\
IC 4040 (CGCG 160252)&$23.0 \pm3.7 $&20.0 &7.81E-14&1.19E-13&1.34E-13\\
NGC 4911 (CGCG 160260)&$25.5 \pm4.2 $&17.5 &5.64E-14&8.23E-14&1.32E-13\\
NGC 4921 (CGCG 160095)&$19.3 \pm4.0 $&17.9 &1.59E-14&3.22E-14&3.54E-14\\
CGCG 160128&$19.5 \pm5.4 $&19.7 &1.27E-14&2.33E-14&2.24E-14\\
CGCG 160127&$24.8 \pm10.3 $&20.0 &1.30E-14&1.86E-14&2.24E-14\\
CGCG 160139&$27.2 \pm4.1 $&21.1 &1.90E-14&2.48E-14&2.98E-14\\
NGC 5033&$19.6 \pm2.7 $&17.6 &1.18E-12&2.34E-12&2.71E-12\\
Mrk 66&$25.0 \pm4.9 $&21.7 &2.76E-14&3.80E-14&4.16E-14\\
NGC 5194&$20.0 \pm3.2 $&18.9 &8.04E-12&1.48E-11&1.55E-11\\
NGC 5236&$21.3 \pm2.5 $&19.7 &1.81E-11&2.99E-11&3.17E-11\\
UGC 8621&$25.0 \pm3.6 $&18.6 &6.79E-14&9.88E-14&1.35E-13\\
NGC 5252&$23.9 \pm3.4 $&20.5 &2.33E-14&3.39E-14&3.82E-14\\
Mrk 266&$28.1 \pm4.4 $&21.6 &3.78E-13&4.81E-13&5.75E-13\\
Mrk 461&$24.5 \pm4.0 $&25.3 &2.01E-14&2.71E-14&2.62E-14\\
Mrk 279&$23.7 \pm3.0 $&20.6 &6.87E-14&1.01E-13&1.12E-13\\
Mrk 799&$27.7 \pm4.2 $&20.1 &5.84E-13&7.59E-13&9.90E-13\\
IC 4397&$27.0 \pm4.0 $&21.6 &5.51E-14&6.94E-14&8.34E-14\\
NGC 5548&$28.5 \pm4.6 $&19.4 &9.10E-14&1.22E-13&1.65E-13\\
Mrk 817&$33.9 \pm6.1 $&24.8 &9.75E-14&1.13E-13&1.28E-13\\
NGC 5719&$27.1 \pm4.0 $&19.4 &4.77E-13&6.37E-13&8.70E-13\\
NGC 5860&$26.5 \pm4.4 $&20.2 &9.14E-14&1.23E-13&1.54E-13\\
Arp 220&$32.3 \pm5.5 $&24.8 &4.79E-12&5.64E-12&6.32E-12\\
NGC 6090&$24.9 \pm4.0 $&22.6 &3.29E-13&4.50E-13&4.73E-13\\
NGC 6104&$22.5 \pm2.8 $&17.0 &3.82E-14&6.54E-14&9.97E-14\\
NGC 6240&$29.3 \pm4.7 $&23.4 &1.09E-12&1.34E-12&1.51E-12\\
Tol 1924-416&$31.4 \pm7.4 $&33.5 &6.77E-14&7.96E-14&7.85E-14\\
NGC 6918&$32.5 \pm5.7 $&21.9 &4.75E-13&5.59E-13&7.09E-13\\
NGC 6946&$18.8 \pm2.8 $&19.6 &9.62E-12&1.87E-11&1.72E-11\\
NGC 7331&$19.2 \pm2.9 $&18.0 &2.91E-12&5.86E-12&6.29E-12\\
UGC 12138&$36.1 \pm9.0 $&22.2 &4.53E-14&5.26E-14&6.65E-14\\
Mrk 323&$23.3 \pm2.9 $&18.5 &2.02E-13&3.16E-13&4.08E-13\\
NGC 7673&$23.2 \pm3.6 $&22.2 &2.47E-13&3.58E-13&3.62E-13\\
Mrk 533&$25.4 \pm5.0 $&21.9 &2.84E-13&3.86E-13&4.24E-13\\
Mrk 534&$28.9 \pm4.5 $&21.9 &3.71E-13&4.63E-13&5.54E-13\\
Mrk 538&$30.8 \pm5.2 $&24.4 &4.82E-13&5.77E-13&6.43E-13\\
Mrk 332&$27.3 \pm4.1 $&19.9 &2.78E-13&3.67E-13&4.82E-13\\
\end{longtable}
\end{document}